# A note on fractional derivative modeling of broadband frequency-dependent absorption: Model III

W. Chen

Simula Research Laboratory, P. O. Box. 134, 1325 Lysaker, Norway

(22 April 2002)

(Project collaborators: A. Bounaim, X. Cai, H. Sverre, A. Tveito, Å. Ødegård)

*Keywords: fractal geometry, fractional derivative, fractional Fourier transform, fractional power of a matrix, self similarity, complex partial differential equation, broadband ultrasound, frequency-dependent attenuation, time domain.*

## 1. Backgrounds

The rational behind this model is schematically illustrated below:

**Fractal geometry** (irregular soft tissues) → **Fractional Fourier transform** (frequency-dependent attenuation: $\alpha(\omega) = \alpha_0 \omega^y$, $y \in [0,2]$ is real valued) → **Fractional derivative** ( Fourier transform $FT_+\left(\frac{\partial^s p(t)}{\partial t^s}\right) = (-j\omega)^s P(\omega)$, $s$ is real valued) → Fractional power of the positive discretization matrix of Laplacian (the modified mode superposition model, see Model I) → **Macro damping effect** ( negative real part of frequency domain solution: absorption)

One definition of the fractional derivative in time is

$$D^s\{p(t)\} = \frac{1}{\Gamma(1-s)} \frac{d}{dt} \int_0^t \frac{p(\tau)}{(t-\tau)^s} d\tau, \qquad 0<s<1$$

where $\Gamma$ is the gamma function. There exist the other mathematical definitions.

**Fractal geometry**: very complicated structures can be described by fractal self-similarity geometry.

**Self similarity:** "An object is said to be self-similar if it looks "roughly" the same on any scale. Fractals are a particularly interesting class of self-similar objects. Self-similar objects with parameters $N$ and $s$ are described by a power law such as

$$N = s^n,$$

where

$$n = \frac{\ln N}{\ln s}$$

is the "dimension" of the scaling law, known as the Hausdorff dimension" (for details see, http://mathworld.wolfram.com/Self-Similarity.html). For power law attenuation, it should be called frequency dimension.

Thus, y in the power law may be a viscous indicator of the cancer tissue, compared with the stiffness (density), since these media parameters have underlying relationships.

It is stressed that the present mathematical modelling is to **represent damping** but does **not necessarily describe the whole physical and chemical mechanisms of damping**.

## 2. Motivation

1). By far, the fractional derivative model is mainly related to the modelling of complicated solid viscoelastic material. **As far as I know, No reference has used this kind of model for medical ultrasound**. Biomaterials are typically complicated with the fractal features of local similarity.

2). Literature concentrates on the **temporal fractional derivative model**. In fact, since the micro geometry of soft tissues which establish the complicated viscous behavior mostly has the fractal dimension structure in space. So, we should instead consider the **spatial fractional derivative model**.

## 3. Linear and nonlinear models

To respect the principle of causality, the damping term should be positive. This leads to the **absolute value of fractional operator (or complex fractional operator)** and the otherwise convolution operation (relaxation). The computing effort of the relaxation model is not trivial. It is noted that for even order operator of damping (independent or squarely dependent of frequency), there is no such absolute value or complex operator issue.

All models given below are fully consistent with the frequency dependent attenuation of any excitations. In particular, we focus on **combining our models with empirical frequency-dependent power law formula** ($\alpha_0$ and $y$ from experiments, $E=E_0 e^{-\alpha(\omega)x}$):

$$\alpha(\omega) = \alpha_0 \omega^y, \ y \in [0,2],$$

where **$y=1$ is most frequently taking place.**

## 3.1. Linear models

**New temporal fractional derivative model** (derived from damped wave eq. y=0):

$$\nabla^2 p = \frac{1}{c^2}\frac{\partial^2 p}{\partial t^2} + \frac{2\alpha_0}{c^{1+2y}}\frac{\partial}{\partial t}\left|\frac{\partial^y p}{\partial t^y}\right| \qquad \text{(real domain)}$$

or

$$\nabla^2 p = \frac{1}{c^2}\frac{\partial^2 p}{\partial t^2} + i^{-3y}\frac{2\alpha_0}{c^{1+2y}}\frac{\partial^{1+y} p}{\partial t^{1+y}}. \qquad \text{(complex domain)}$$

**New spatial fractional partial derivative model** (derived from augmented wave eq. y=2):

$$\nabla^2 p = \frac{1}{c^2}\frac{\partial^2 p}{\partial t^2} + \frac{2\alpha_0}{c^{1+y}}\frac{\partial}{\partial t}\left|\nabla^y \bullet p\right| \qquad \text{(real domain)}$$

or

$$\nabla^2 p = \frac{1}{c^2}\frac{\partial^2 p}{\partial t^2} + i^{-3y}\frac{2\alpha_0}{c^{1+y}}\frac{\partial}{\partial t}\left(\nabla^y \bullet p\right). \qquad \text{(complex domain)}$$

## 3.2. Nonlinear models

Unlike the preceding hyperbolic PDE models, Burgers equation is a parabolic PDE models,

$$p_t + p \bullet \nabla p - \varepsilon \nabla^2 p = 0$$

where the attenuation is due to diffusion modeled by the second order spatial derivative. The model has the attenuation of quadric frequency dependence. In the case of anomalous diffusion (0<y<2), the fractional derivative comes into play.

**New modified Burgers equation** (the standard Burgers eq. y=2)

$$p_t + p \bullet \nabla p + 2\alpha_0 \left| p^{2-y} \left( \nabla^y \bullet p \right) \right| = 0$$

or

$$p_t + p \bullet \nabla p + i^{-3y} 2\alpha_0 p^{2-y} \nabla^y \bullet p = 0.$$

So, in the case independent of frequency (y=0), the modified Burgers equation is

$$p_t + p \bullet \nabla p + 2\alpha_0 p^2 = 0.$$

## 4. Open computational and analysis issues

1) $\alpha_0$ and y may have some connections with density, viscous parameter, wave speed and other physical parameters of soft tissue, which can be derived from degeneration relationship between the above PDE models and the classic models.

2) Fractal dimensionality may have something to do with parameter $\alpha_0$ and y, which coincides the different morphologies of the cancer and normal tissues. The fractal geometry has been used for cancer detection.

3) FEM discretizing the absolute value of spatial operator or complex PDEs? Note the absolute value is to ensure the positive definition of damping fractional derivative. So, if we could find and prove the positive definition with respect to some y, and then, the absolute value operation can be removed. In the later Appendix, we give a FEM scheme via the fractional power of a matrix to calculate the absolute PDEs.

4) I have no idea of FEM discretization of fractional derivative in space. There are plenty of reports on the finite difference approximation of fractional time derivative.

5) The related stability analysis is unclear.

6) Fractal geometry, fractional derivative, what is the missing **fractional algebra**? Does it have something to do with the fractional power of a matrix as proposed in the previous modified superposition model? A systematic research of all these issues is still lacking.

7) Complex modal analysis (frequency domain) has become very important in modal analysis and parameter identification. I wonder if it is indispensable to use complex and fractional derivative describing the behaviors of complicated soft tissues.

**Appendix**

1. FEM formulation for the absolute value of partial differential operation:

$$\left|\nabla^y u\right| = \left(\nabla^2\right)^{y/2} u = A^{y/2} u,$$

where $A$ is the FEM discretization matrix (symmetric positive definition) of the Laplacian. It is worth noting that **the spatial fractional derivative model in real domain** could be understood as the PDE correspondence of the previous **modified mode superposition model**.

In the case $y=1$, $\nabla u = Bu$, is $B$ a skew symmetric matrix? If so, $B \neq A^{1/2}$ since the latter is still symmetric, but $\left|\nabla u\right| = A^{1/2} u$.

2. In time, the situations seem more certain. $\frac{\partial^2 u}{\partial t^2} \prec 0$ and $\frac{\partial u}{\partial t} \succ 0$. Are they right? Then, how about $\frac{\partial^y u}{\partial t^y}$ for the general $y$?

4. Attenuation and dispersion is closely interdependent. We should have a frequency analysis of the above fractional derivative model in time and space.

5. The classical **structural damping** model

$$(i\eta + 1)\nabla^2 u = \frac{1}{c^2} u_{tt} \quad \text{(complex domain)}$$

is independent of frequency, where the damping coefficient $\eta$ is determined via complex frequency analysis. The mechanism is that this damping is proportional to the displacement rather than the velocity. A combination of the structural and viscous damping may occur in some soft tissues.

6. Complex PDEs also appear in the Schroedinger's equation for modeling quantum mechanics problem. So, it is not a fuss to have a complex derivative in our model.

7. After a rough look at the fractional derivative, Szabo's convolution model may be equivalent to the present fractional derivative. It is noted that the Szabo's model is not easy to get a numerical solution. In contrast, the fractional derivative in time has the standard discretization formulation and related numerical analysis.

8. The FEM discretization of partial fractional derivative seems an underdeveloped issue by now. As far as I know, there are much fewer fractional derivative models in space than in time.